\documentclass[]{elsarticle}

\usepackage{amssymb}
\usepackage{graphicx}
\usepackage[table,xcdraw]{xcolor}
\definecolor{lightgray}{gray}{0.9}
\RequirePackage[utf8]{inputenc}
\usepackage{url}
\usepackage{color}
\usepackage{epsfig} 
\usepackage{hhline} 
\usepackage{multirow} 
\usepackage{longtable} 

\newcommand{\vrc}[1]{\textcolor[rgb]{0,0,0}{#1}}

\newcommand{\hide}[1]{}

\usepackage{numcompress}
\usepackage{setspace}

\journal{X}

\begin{document}

\begin{frontmatter}

\title{Experimental Security Analysis of Controller Software in SDNs: A Review}



\author[unifesp]{Tiago V. Ortiz} 
\ead{tvortiz@unifesp.br}
\author[unifesp]{Bruno Y. L. Kimura}
\ead{bruno.kimura@unifesp.br}
\author[icmc]{Jó Ueyama}
\ead{joueyama@icmc.usp.br}
\author[unifesp]{Valério Rosset \fnref{myfootnote}}
\ead{vrosset@unifesp.br}

\address[unifesp]{Instituto de Ciência e Tecnologia, Universidade Federal de São Paulo (UNIFESP), Av. Cesare M. G. Lattes, 1201, 12247-014 - São José dos Campos-SP, Brazil}
\address[icmc]{Instituto de Ciências Matemáticas e de Computação - ICMC, Universidade de São Paulo (USP), Avenida Trabalhador São-carlense, 400, 13566-590 - São Carlos-SP, Brazil.}
\fntext[myfootnote]{Corresponding author.}

\begin{abstract}

The software defined networking paradigm relies on the programmability of the network to automatically perform management and reconfiguration tasks. The result of adopting this programmability feature is twofold: first by designing new solutions and, second, by concurrently making room for the exploitation of new security threats. As a malfunction in the controller software may lead to a collapse of the network, assessing the security of solutions before their deployment, is a major concern in SDNs. In light of this, we have conducted a comprehensive review of the literature on the experimental security analysis of the control plane in SDNs, with an emphasis on vulnerabilities of the controller software. Additionally, we have introduced a taxonomy of the techniques found in the literature with regard to the experimental security analysis of SDN controller software. Furthermore, a comparative study has been carried out of existing experimental approaches considering the security requirements defined by the Open Network Foundation (ONF). As a result, we highlighted that there is a need for a standardization of the methodologies employed for automated security analysis, that can meet the appropriate requirements, and support the development of reliable and secure software for SDNs.

\end{abstract}


\end{frontmatter}



\section{Introduction}\label{sec:introduction}
The growing demand for the  efficient  management and reconfiguration of large-scale communication services in computer networks, has leveraged the adoption of the network programmability paradigm of the Software Defined Networks (SDNs)~\cite{I16}. The flexibility provided by  network programmability allows several solutions to be deployed for complex tasks and thus reduces the need to replace the hardware/firmware of  network devices. Among the many advantages of using SDNs, is that it enables autonomic services to be installed which are aimed at the optimization of network metrics, network virtualization, migration, mobility and energy conservation \cite{I8}.


Unlike the case of traditional network architecture, in a SDN, the switches comprise  the data plane and have two key  features. First, the switches interact with a logical network control entity, known as the SDN controller, to obtain forwarding instructions. Second, they keep a flow table in which entries with instructions are stored and evaluated so that they can  carry out the forwarding of packets~\cite{I63}. Generally, the function of the SDN controller is operated  by  non-standard software that provides the necessary  interfaces  for network services and management. In SDNs, the interaction between switches and controllers generally occurs through an application programming interface (API). The OpenFlow (OF) ~\cite{mcKeown2008} protocol is recognized as the standard API for SDN and it is  recommended by the guidelines of the {Open Network Foundation} (ONF)~\footnote{www.opennetworking.org}. Hence, the applications hosted by the controller can implement policies, routing algorithms and other management services (e.g. load balancing, virtualization, support for mobility etc.) to determine a set of entries in the flow tables of the switches. Thus, the controller and applications act as the ``brain" of the network by forming the logical network control.


Security in SDN is a major concern that has recently attracted the attention of the scientific community \cite{shu2016, Abdou2018}. This can be explained by the fact that  the programmability and centralized management of SDNs may either assist in  the implementation of security services or lead to the emergence of new security threats which may compromise data and the network operations~\cite{I60}. Comparatively, SDNs are more vulnerable to attacks  than legacy networks \cite{shu2016}. It is noteworthy that the OF specifications stipulate  that communication between the controller(s) and switch(es) must occur through a secure channel, that is, it must be  implemented by secure transport services (e.g. TLS). As well as this, owing to the fact that many interfaces/operations in SDNs are not standardized, there is  no guarantee  that any single mechanism can ensure proper security.


For this reason, the scientific community has dedicated a great deal of  effort to   investigating potential security risks in SDNs \cite{dacier2017}. On the one hand, one may observe  in the related literature the existence of several papers concerned with i)  carrying out  reviews of security in SDNs, ii)  defining concepts in  theoretical terms  and iii)  discussing  possible security vulnerabilities, threats and countermeasures \cite{Scott-Hayward2013,I60,I7,Li2016,Dargahi2017}. On the other hand, an attempt has been made to conduct experimental security analysis with focus on the detection of vulnerabilities in different  planes of the SDN architecture. In addition, another important factor that should be taken into account is the set of security requirements applicable to each SDN plane. In specific terms, in the control plane context, \cite{TR-529} describe the security requirements for SDN controllers, that are derived from an analysis and classification of threats defined by the ONF in \cite{TR-530}. However, as there is a great heterogeneity of methods designed for experimental security evaluation in the literature, it is necessary to understand which of them can be applied to assess the fulfillment of the recommended security requirements by the controller software implementations.


\vrc{In view of this, we present in this paper the results of an extensive review of the most recent literature dedicated to the experimental security analysis of controller software in SDNs. Our research study complements the findings of current surveys that focus on general aspects of SDN security, e.g. \cite{shu2016,I60,E5,A2,I63}, by providing a broader discussion of the techniques employed in the literature for the experimental security analysis of controller software.} In addition, we introduce a taxonomy of the approaches found in the literature review and conduct a comparative analysis of them considering the security requirements defined by the ONF and the STRIDE security threat categorization model \cite{P2}.

The remainder of this paper is structured  as follows.


In Section \ref{sec:SDN} we describe the theoretical concepts  of SDNs with focus on its architecture,  main components and security threats. In Section \ref{sec:review}, we carry out  an extensive literature review with regard to the security analysis of controllers and other closely-  related components of SDNs. A discussion about the data collected during the literature review is presented in Section \ref{sec:results_review}. Following this,  we introduce a taxonomy and a discussion of the main methods employed for  the experimental security evaluation of SDN controllers in Section \ref{sec:methods} and Section \ref{sec:discussion}, respectively. Finally, we summarize the conclusions of  the paper in Section \ref{sec:conclusion}.

\section{SDN: Architecture and Security Threats}
\label{sec:SDN}

\begin{figure*}[!ht]
\includegraphics[height=0.70\textwidth]{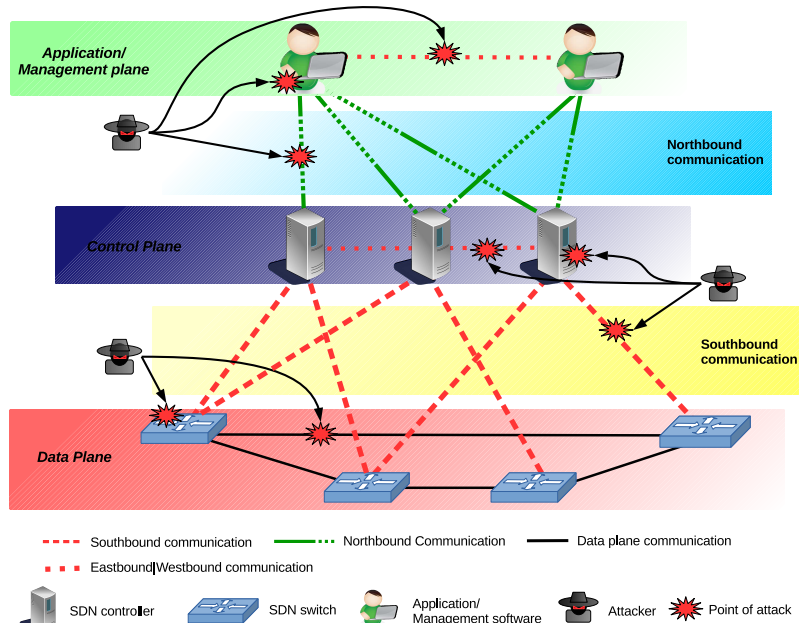}
\caption{SDN architecture abstraction.}
\label{fig:sdn_general}
\end{figure*}

The purpose of the paradigm adopted by the SDNs is  to decouple logical network control from  packet forward by dividing the networking tasks into three planes: the data plane, the control plane and the application/management plane \cite{I63}. Figure ~\ref{fig:sdn_general} shows an overview of the SDN architecture with  the three planes included. In the SDN architecture, the interaction between the network components in each plane occurs through specific interfaces. In this case, each interface is classified as belonging to one of the four extremities of the architecture: northbound, southbound, westbound and eastbound \cite{I8}.


The data plane essentially consists of network switches that forward frames,  belonging to each flow, to other devices by means of  a set of locally- installed flow rules. The SDN controller is able to  configure the forward task by installing the flow rules into the switches proactively or reactively. In the reactive forwarding, on receiving the first frame of a flow, a switch usually requests  the SDN controller to lay down  the rules regarding that individual flow, and to keep  those rules active until the end of each stream. In the proactive forwarding, the controller sends  a set of rules to the switches that are defined \textit{a priori} for each flow. In the absence of a rule for a specific received flow, the switch may query the controller to obtain the new rule or forward the flow in the traditional way in accordance with  a prior configuration of the device.

In the control plane, the controller acts as a network operating system (NOS) and provides  the basic abstractions and functionalities needed for the implementation of the forward policies defined by the management entity of an Autonomous System (AS). Controllers can  communicate with  each other  to ensure  distributed network control. Alternatively, a controller can  interact either with other controllers belonging to a different  AS or with legacy devices by means of the westbound and eastbound interfaces, respectively \cite{hakiri2014}. In addition, a controller can interact  with the data plane through southbound protocols. In this particular situation, although there exist  several APIs and protocols designed for SDNs, (such as the Forwarding and Control Element Separation (ForCES) \cite{doria2010}, the Open vSwitch Database Management Protocol (OVSDB)\cite{pfaff2013} and the Protocol-oblivious forwarding (POF)\cite{song2013}), the OpenFlow protocol emerged as a \textit{de facto} standard owing  to its wide acceptance by practitioners and manufacturers \cite{I63}.


The application/management plane allows the network management entity to monitor and execute control operations on the SDN, usually by interacting with the controller(s) via northbound APIs/protocols, (e.g. RESTFUL APIs or programming languages \cite {I63}). It should be  mentioned that, although the theoretical SDN architecture suggests that  the application and control planes software are completely independent, this premise is rejected by the current state of SDN controller software implementations. Generally, both the control and application are a part of the same software and share computational resources. In light of this, the northbound interface is often  used by the applications for communication with external entities instead of the controller. Hence , as applications may operate as plug-ins or particular  modules tied to the controller software, it is not unusual to experience  collateral damage  to the controller software caused by the  malfunction of some application.


Security in SDNs is an important research field  because the programmability and centralized management of SDNs may either assist in  the implementation of security services or lead to the emergence of new security threats which , in turn, can compromise data and the network operation \cite{I60}. Comparatively, the SDNs are  more vulnerable to  attacks  than legacy networks \cite{shu2016}. In this sense one may enumerate at least seven possible points of attack in SDNs (as shown  in Figure~\ref{fig:sdn_general}), which are as follows: the SDN switch, the data communication between switches, the SDN controller, the southbound communication between controller(s) and the switch(es), the east/westbound communication between controllers or between application/management software, the northbound communication between the controller and the application/management software and the management/application software. It should be noted  that the OpenFlow specifications  state  that communication between the SDN controller(s) and the switch(es) must occur through a secure channel (e.g. by using TLS). Despite this, largely owing  to a  lack of standardization, there are no guarantees that the current  mechanisms are suitable  and can meet  all the security requirements of SDNs. 


Keeping these points in mind, in the next section, we outline  the state-of-the-art with regard to the experimental security analysis of controller software in SDNs. 

\section{Literature Review}
\label{sec:review}


For this review, we carried out a rigorous and extensive verification of the literature by checking  the bibliographic repositories available on the Internet. We selected a set of papers considering only the most relevant studies \textbf{fully or partially dedicated to the experimental security analysis of the controller software in SDNs}. More specifically, the literature review mainly includes studies  related to the following controllers: OpenDayLight, Floodlight, Beacon, Ryu, POX, ONOS, NOX, Maestro and their variants.


\vrc{Whilst there exist multiple solutions for south-bound SDN interfaces, e.g., PCEP~\cite{vasseur2009path}, NETCONF~\cite{enns2006netconf}, P4~\cite{bosshart2014p4}, ForCES~\cite{doria2010}, I2RS~\cite{hares2013software}, the OpenFlow~\cite{mcKeown2008} is by far the most widespread existing solution. Thus, by examining the most recent  experimental  analyses of security issues with regard to  open SDN controllers in the literature, we necessarily report studies related to innate OpenFlow-based SDNs. For further  detail  about the security analysis of other south-bound protocols in SDNs, the reader should  refer to \cite{brandt2014security}.}

\begin{figure}[!ht]
\centering
\includegraphics[height=0.60\textwidth]{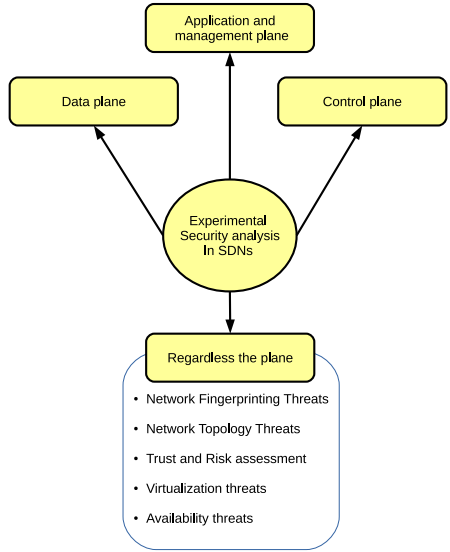}
\caption{Experimental security analysis of SDN Controllers road-map.}
\label{fig:sdn_literature}
\end{figure}

Among the set of analyzed papers presenting practical and theoretical results, we selected only the experimental studies to be described in the following subsections. For ease of understanding, the selected papers were grouped according to the main focus of each experiment carried out, as depicted in Figure \ref{fig:sdn_literature}, in: data plane; application and management plane; control plane and regardless the SDN plane. \vrc{After describing the selected experimental studies in this section, we provide a comprehensive analysis of their security aspects in Section~\ref{sec:results_review} and Section~\ref{sec:methods}.}

\subsection{Security Analysis in the Data Plane}


In this section we describe the experimental security analysis with regard to the data plane, presented in \cite{I17}, \cite{I32} and \cite{Nguyen2017}, respectively.


The authors of \cite{I17} analyzed the effects of overhead on  the routing tables of the SDN switches. As an experiment, the authors implemented a scenario in which an attack was  carried out by continuously sending several flows of forged raw packets. In this case, only a small portion of the packet header was  changed between streams, so that an individual entry could be  inserted into the switch for each forged packet until the storage capacity was  exhausted. According to the authors of \cite{I17}, when  the resources of the switch are exhausted, the controller is unable to install new entries for legitimate flows. To mitigate this type of attack, they recommend  that the controller should  either wait until a flow entry expires before  inserting a new entry of a legitimate flow or directly forward the legitimate flow through to  another switch. 
The authors also conducted  tests to measure the delay,  throughput, and  packet drop rate when the  flow table is exhausted. The measurements were performed on the {OpenDayLight} (ODL) controller that was configured with the two mitigation strategies. The tests considered legitimate flows with only a subset of entries already installed in the switch and forged flows with no related entries in the switch tables. As a result, the evaluation detected  a performance degradation in  the controller. In addition, the tests also revealed a degradation in the throughput of the legitimate flows when there were  no entries installed in the switch; this was  mainly due to the extra work carried out  by the controller to forward the packets to other switches.


The authors of \cite{I32} introduced a prevention scheme against attacks originating in compromised OpenFlow switches. A system scenario is defined in which a compromised switch is able to launch three types of active attacks, namely, incorrect forwarding (deviation or duplication of a flow), packet manipulating and a malicious weight adjustment  of the compromised switch group tables. The authors created two algorithms   to detect these attack vectors. The purpose of the  first algorithm, called \textit{Forwarding Detection}, is  to verify  if the flows are being correctly routed by the switches. To do this , the algorithm uses an artificial packet to check the flow entries in the switch tables and tracks the path traveled by the flow to determine the correctness of the path. The second algorithm, called   \textit{Weighting Detection}, is used to check if the switch forwards packets at an expected rate. The algorithm generates and sends a set of artificial packets to the evaluated switches and analyses the amount of received packets. If the number of received packets is close to the expected value,  the flow is considered to be satisfactory; otherwise the flow is  compromised.


Recently, in \cite{Nguyen2017} the authors evaluated the link discovery service (LDS) vulnerability, which is an essential  service of the control plane that is needed for the proper operation of SDN services and applications. The authors provided  details of how an attacker can "poison" the perception of the SDN controller about the network topology, by introducing false links. They also explained that such links can be created through the use of false Link Layer Discovery Protocol (LLDP) messages, sent by one or more compromised hosts. As a result, the authors were able to simulate a scenario of a realistic link falsification attack on the FloodLight controller implemented on the Mininet emulator.

\subsection{Security Analysis in the Application Plane}


In this section we describe different approaches in the literature committed to assess experimentally the security of the application plane and the interfaces for communication with the control plane. 

The study \cite{A13} discusses the  application failures that can  result in the crash of the SDN controller.  More specifically, the authors address the problem of fail-stop crash failures  and Byzantine faults. The authors claim that the availability of a SDN controller is decreased by the sharing of resources between the SDN applications and controllers. This means that  the crash of the former induces the crash of the latter, and hence  affects  availability. They  also examine  the relationship between applications and the network, where  a Byzantine failure may lead to the violation of  network safety property and thus affect  network availability.
In view of this, the authors  re-design  the SDN controller architecture by focusing on a set of abstractions to increase the resilience of controllers to faults caused by applications. This new   design has two objectives: to isolate the applications of the controller in order to eliminate resource sharing and isolate the network from application to allow  consistent  support for  applications. The authors  tested and  validated the new architecture by  creating  a  FloodLight-based isolation system called  LegoSDN \vrc{and   discussed  the hypothetical challenges of their prototype}.


The authors of \cite{A6} argue that application failures may either lead to losses in the control plane or cause the permanent failure of the network operations. They claim that problems caused by application failures directly interfere with the proper functioning of the SDN controllers. Moreover, they indicate a lack of solutions for this problem in the existing controllers. For this reason, they developed a new controller called {Rosemary}, in order to offer more resilience to application faults. \vrc{The main features  of {Rosemary} are a) its ability to   isolate  the controller from applications, b) it can  spawn   applications as independent processes, and also c) to monitor and  control the resources of each application.} The authors compare {Rosemary} with three other  controllers, namely, NOX, Beacon and FloodLight. In the experimental analysis, they performed crash tests of controllers induced by applications, memory leakage tests in the controllers and unauthorized access test of  memory structures. The NOX, Beacon and FloodLight controllers failed  the tests, while Rosemary proved to be resistant to the attacks \vrc{due to the isolation provided by its  architecture}. Finally, the authors carried out a performance comparison, which took account of  the flow rate, where the performance of Rosemary was  far superior to that of FloodLight and NOX but, in general, was similar to that of Beacon.


In \cite{A10}, the authors introduced attacks in the application plane in which an adversary spreads  malicious applications to compromise the operation of the network and also discussed possible protective  mechanisms  against such attacks. They carried out  three case studies of attacks that were triggered by  malicious applications installed in  the FloodLight, ONOS, and OpenDayLight controllers. \vrc{The authors argue that a permission- checking mechanism could be implemented to protect against these  attacks. Moreover, in addition to this  mechanism, they also suggest conducting a static source code analysis and a dynamic analysis based on code coverage test of the target application. }

\subsection{Security Analysis in the Control Plane}
\label{sec:TrabCont}


Here we describe the approaches employed for assessing security in the control plane and comparing the communication interfaces with the other planes in SDNs. 


The authors of \cite{I31} investigate the impact of software aging (SA) on SDN controllers. The concept of SA was used to analyze the performance and security of a system by observing the operating conditions of a software over an extended period of time. The authors  analyzed the impact of the SA on the FloodLight and Beacon controllers. In the experiments, the controllers remained in operation for $36$ hours, with alternating periods of high workload and idleness  lasting for  twenty and five minutes, respectively. \vrc{The authors used the {cbench} tool to generate the workload (packet-in flows) for $16$ Openflow switches at a  rate of $100.000$ flows/s.} As a result, it was found  that FloodLight used all the available resident memory ($4.2$GB) of the Java Virtual Machine (JVM) as well as all the virtual memory, including that stored on disk, making a total of $8$GB of memory consumption. The memory management of Beacon was better since it used approximately $500$MB of resident memory and $5$GB of virtual memory.


The study carried out by  \cite{I44} is concerned with  unauthorized access to SDN controllers including  threats originating from applications. The authors propose a set of  Authorization and  Access  Control rules as well as a reference monitor named \textit{OperationCheckpoint}, which is coupled to the FloodLight controller. The  improvements  include the following:  a) the definition of a set of rules for the OpenFlow commands of a given application, b) the creation of unique identifiers (IDs) for  applications, c) Permission  Management  and d)  persistent records of actions and access to the controller. The authors  validated the scheme by  evaluating  the \textit{OperationCheckpoint} in an environment emulated with {Mininet}. The experiments were divided into two scenarios: one considering access to the REST API and another considering access to the internal Java methods implemented in the controller. The proposed \textit{OperationCheckpoint} prevented access to calls for the REST API methods, but allowed  calls made through Java methods due to the internality of Java. Finally, the authors measured the latency that was introduced by using the \textit{OperationCheckpoint}. There was an  average increase of approximately $367.125\mu$s  in the latency of the operations performed by the  controller.


Similarly to \cite{I44}, the authors of \cite{I47} designed  a configurable mechanism to restrict access to system-level operations through the application of security policies. The purpose of the system  is to maintain control over third-party applications installed in  the SDN controller and thus  prevent attacks resulting from malicious actions. In the mechanism,  each network service that can carry out  sensitive operations, is held in a \textit{sandbox}. Two modes of operation were defined: detection and protection. In the detection mode, the operations are monitored and the security breaches are only recorded in the log file. In the protection mode, \textit{sandboxes} can restrict the execution of operations by complying with  the security rules defined by the network administrator. The mechanism was implemented and validated in the OpenDaylight controller within  the Mininet environment. The authors validated the  mechanism by conducting  effectiveness and performance tests. The purpose of the effectiveness test was  to determine whether the  security rules were being enforced for   the strict execution of operations from two different  applications. It was found that the  mechanism was able to contain  the malicious operations satisfactorily in the tests. As a result of the performance test, it was verified a reduction of less than $5$\% in the effective transmission rate of the network  due the control overload introduced by the proposed mechanism.


The authors of \cite{A1} investigated the vulnerabilities in the OpenDayLight controller with regard to   Man-in-the-Middle (MitM) attacks. The study addresses two variants of threat vectors, namely, threats in the communication of the control plane and threats in the controller. The authors presented a MitM attack, which intercepted  messages between the SDN controller and a remote client, with the aim of  capturing the   credentials of the controller. When carrying out the attack, the authors assumed a scenario where the adversary is located on the same LAN as the SDN controller and can capture packets. For this reason, they executed the MitM attack on the communication of a remote terminal and the OpenDayLight controller via the web DLUX interface of the controller. In the experiment, the adversary captured HTTP (non-SSL) messages exchanged between the {web DLUX} interface and the remote terminal. As a result, the credentials of the controller were captured in plain text.


\cite{A2} conducted  an experimental performance analysis of the respective open source controllers: NOX, POX, Beacon, FloodLight, MuL, Maestro and Ryu. The selected controllers were compared in terms of  performance, scalability, reliability and security metrics. The flow rate and latency metrics were used to  measure performance and scalability  and covered  scenarios in which  the number of hosts ranged from $1$ to $256$ and the number of switches from $10^3$ to $10^7$.  In the case of   the reliability analysis, the authors measured the number of connection failures and packet losses that  occurred during an operational period of $24$ hours by examining  workload values between $2,000$ and $18,000$ requests per second. Finally, when investigating  security issues, five types of malformed OpenFlow packets were created to observe the behavior of the controller on  receiving these  packets. As result of the experiments, it was noted  that the Beacon controller had a  higher flow rate while the POX, NOX and Ryu had a very low flow rate. In addition,  Mul and Beacon had  the lowest latencies. With regard to  reliability, Mul and Maestro presented higher packet drop rate than  the other controllers. Finally, despite having a poor performance, RYU achieved  the best security classification level.


The authors of \cite{A12} designed  a new extension for  the NOX controller, called  {FortNOX}, which implements a role-based authorization strategy, security restrictions and conflicts resolution between flow rules.  FortXOX enables  NOX  to check for real-time flow rule conflict. FortNOX employs  a rule-based representation method, called alias reduced rules (ARRs), and detects  conflicts  by comparing the set of ARRs and the new installed rules. The authors  validated the scheme by estimating   the computational time aggregated by the conflict analysis and resolution employed in the FortNOX. As a result, the FortNOX conflict analysis showed , on average, an increase of $7$ms in the processing time for each new flow rule request.


The authors of \cite{IS2} introduced a security extension to the control plane to provide security management and arbitration of conflicts which  originated from the insertion  of multiple flow rules by distinct applications. The proposed extension layer, called the Security Enforcement Kernel (SEK), was implemented on top of the FloodLight controller. The SEK conducts  a Rule-based Conflict Analysis (RCA) which acts as a  conflict resolution mechanism that mediates all the requests for installing   new flow rules. The SEK has the following components: an code authentication module; conflict detection  and  resolution; a state manager; callback tracking; a permitting mediator; a flow policy synchronizer; the separation of processes and a security audit service. 

 
\cite{IS1} present several attacks on controllers that violate the network topology and the forwarding strategy in the data plane of SDNs. In addition, it  demonstrated the feasibility of launching attacks on four controllers: Maestro, POX, OpenDayLight and FloodLight. For this reason, the authors proposed a new controller, -  SPHINX. This  controller is capable of detecting attacks by dynamically learning new network behavior  and giving an  alert when it detects a suspicious activity. In the experiments, the authors evaluate the controllers with regard to  the following attacks: ARP poisoning, Fake topology, Controller DoS, Network DoS, TCAM exhaustion and Switch blackhole. As a result, they observed the lack of resilience of OpenDayLight, FloodLight and POX to TCAM exhaustion, DoS and Fake topology attacks, respectively. They report the resilience of SPHINX to all the attacks carried out. Additionally, with respect to performance, the results show that SPHINX can verify $1,000$ policies at each network update in $\approx 245 \mu$s, by generating an increase of $\approx 6$\% and  $\approx14.5$\% in the processing and memory usage, respectively.

\subsection{Security Analysis Regardless the Plane}


In this section, we report the experimental approaches concerned with the evaluation of the security without taking account of any  specific plane in  the SDN architecture. For purposes of clarification , the  approaches are grouped in accordance with their research objectives, as follows  : Network fingerprinting threats, network topology threats, trust and risk assessment, virtualization threats and availability threats.

\subsubsection{Network Fingerprinting Threats}


The authors of \cite{I28}, \cite{I41} and \cite{A3} address the problem of the fingerprinting of controller-switch interactions in SDNs. By fingerprinting the SDN, an adversary can estimate the packet-forwarding logic, map the network topology and identify controllers, which are all factors that  may expose  the network to threats. 


In \cite{I28} and \cite{I41}, the authors investigated the ability of a remote adversary to determine (passively or actively) if any flow rule installation has been  triggered by a given packet. The rule of recognition attack is carried out  by releasing  new streams with a cross-traffic generator. The results are obtained by means of the statistical analysis of the Round-Trip Time (RTT) and packet-pair dispersion of these flows. The dispersion between packet pairs refers to the time interval between the complete transmission of two packets sent by a client on a particular link \cite{I28}. When conducting  the statistical analysis, the authors consider two scenarios: in the first scenario, the packet does not trigger a new installation of rules, while in the second scenario the packet does. In the experiments, it was possible to determine that  a packet is able to  trigger  the installation of a new flow with an accuracy of $98.54$\%.

In \cite{A3}, authors devised  a network scanning tool called \textit{SDN scanner} for remotely fingerprinting a SDN network. The \textit{SDN scanner} sends fake packets to a particular  target network and repeatedly estimates  the response time of each packet. In this way, the tool can determine whether  flows are  new or already  exist. Subsequently, the scanner statistically tests the samples from the two sets to determine whether a given network is SDN or not. This can be carried out  by means of statistical tests, such as \textit{t-test}\vrc{~\footnote{\vrc{The \textit{t-test} is a well-known statistical hypothesis testing method that can be used to determine if two sets of data are significantly different from each other.}}}. As a result of the experiments, the tool achieved a  fingerprinting rate of $85.7$\%.

\subsubsection{Network Topology Threats}


The authors of \cite{I43} evaluated the security in SDNs with a special focus on the {OpenFlow} protocol. In their study, the {OpenFlow} is analyzed on the basis of  the STRIDE security threat categorization model. When conducting  the analysis, a data flow diagram (DFD ) is used to model the OpenFlow so that it can be combined with the vulnerabilities of the STRIDE threat categorization model, and thus  result  in several Attack Trees. After  the trees have been  generated, practical tests are carried out  to validate the  approach, such as: Denial of Service (DoS) in the switch flow table, Denial of Service (DoS) in the switch input buffer and fingerprinting the network to determine if a rule exists on a switch and if  the activity of a particular client can be tracked.


\cite{I66} assessed  the security of the Open Flow Discovery Protocol (OFDP), a variation of the Link Layer Discovery Protocol (LLDP). The OFDP is part of the OpenFlow software distribution. However, OFDP does not support authentication, integrity or confidentiality. Thus , it is susceptible to network topology poisoning attacks through spoofed links. The authors of \cite{I66} conducted experiments in which it was possible to create unidirectional and bidirectional false links by using one and two infected hosts, respectively. In another scenario consisting of five switches and five client hosts, the network connectivity is reduced by $30$\% through  the creation of one false link. As an improvement, the authors suggest the use of a Keyed-Hash Message Authentication Code (HMAC) to ensure authenticity and integrity in the network. The authors report that the use of OFDP with HMAC increases the network processing load by approximately $8$\%.


The authors of \cite{IS3} examine  attack vectors for  SDNs based on the exploitation of false links. The attacks are intended to poison the topology of the network to mislead the core services of the controllers and applications. The authors found failures in the Host Tracking Service and Link Discovery Service systems in the SDN controllers. This led to a pair of  poisoning attack  groups in the network, namely Host Location Hijacking and Link Fabrication attacks. The Host Location Hijacking attack group encompasses Exploitation in Host Tracking Service (HTS) attacks, in which an adversary impersonates a legitimate host to capture the network traffic destined for the legitimate host, and the Web Clients Harvesting, where the attacker impersonates a legitimate web server. The Link Fabrication attack includes five variations: Exploitation in Link Discovery Service, Fake LLDP Injection, LLDP Relay, Denial of Service Attack, and MitM. The proposed attacks were successfully carried out  on the FloodLight, OpenDayLight, Beacon and POX controllers. As an improvement, the authors added  a new security extension to the Floodlight controller, called  TopoGuard, which provides automatic and real-time detection of Network Topology Poisoning Attacks.  


The study \cite{E4} analyses the impact of attacks to SDNs through the latency and packet loss observed in web services traffic. \vrc{Without pointing out possible countermeasures, the authors  discuss  vulnerabilities and security threats, by highlighting the following : threats to SDN management, threats to the control plane and threats to the data plane.} Two types of attacks were carried out. In the first attack, an adversary sends false requests as a means of keeping  the controllers and switches busy, as well as causing delay and data loss. In the second attack, a malicious switch is employed  to impersonate a legitimate switch, which leads to  a loss  of connection with the SDN controller.

\subsubsection{Trust and Risk Assessment}


The study conducted by \cite{I70} deals with the question of  trust analysis in SDN controllers and their applications. The introduced approach relies on several redundant controllers that are  capable of operating in different  environments. The authors introduce  an intermediate SDN layer that is  located between the switches and the controllers, called Trust-oriented Controller Proxy (ToCP), and based on a hypervisor. The ToCP layer collects and analyses the rule installation requests originating from  distinct controllers and if they are  deemed consistent and trustworthy, it installs the rules on the network switches. \vrc{The authors made  an evaluation of ToCP in a SDN network by taking note of  the throughput, packet loss, Jitter, memory and processing overheads. More specifically, the functionalities of ToCP increased the Jitter and packet loss and reduced  the throughput by $13.6$\% in the worst case scenario. Similarly, the memory consumption and CPU usage were also increased by $401$MB and $0.4$\%, respectively.}


In \cite{S1}, the authors classified  security threats by introducing a method of risk assessment and countermeasures for some of the security issues pointed out. They drew up  a security checksheet to  help  network designers to determine risks and find suitable countermeasures to mitigate them. The list of attacks was  combined with a \textit{prior} list described in \cite{A11}. As an experiment, the authors first  made a  qualitative assessment of an  SDN testbed by using the proposed checksheet. Afterwards, two Denial of Service (DoS) attacks were carried out  to quantitatively measure the performance degradation of the network.

\subsubsection{Virtualization Threats} 


The authors of \cite{S15} analyzed the FlowVisor, which allows  network virtualization in SDNs. They  found vulnerabilities in the FlowVisor isolation mechanism, which were at risk of  being exploited by a malicious entity in the network. Moreover, the  vulnerabilities that were found, allow traffic manipulation within different  virtual networks. The authors state  that FlowVisor does not verify permissions of SDN controllers. This allows a malicious controller to send packets to other controllers through FlowVisor. Thus, three variations of attacks were identified: the VLAN ID access problem, field modification problem and the wild card modification problem.

\subsubsection{Availability threats}


It is noteworthy that a great number of papers in the literature  address the availability of resources by exploring Denial of Service (DoS) attacks in SDNs. Among them, we selected one study to represent this category in this  review, \cite{alharbi2017}. A  comprehensive review of Distributed DoS attacks on  SDNs can be found in \cite{Yan2016}. 


In \cite{alharbi2017}, the authors carry out  an experimental evaluation of the impact of DoS attacks on SDNs by looking at  two types of DoS attacks against the control and data planes. In the data plane, they examine  two types of attacks aimed at  exhausting both the resources of the SDN controller and switches.  The authors concluded  that an adversary controlling a single host is able to  disrupt the forwarding capability of a SDN network with relatively limited resources.

\section{Experimental Environments for SDN Security Analysis}
\label{sec:results_review}


In this section, we present a comparative analysis of the literature. In the case of  each research study analyzed, we examined  the following features of the experiments conducted by the authors: the controllers used, the environments  and their hardware and software configurations and the supporting tools for the security analysis.

\subsection{Main Controllers}
\label{sec:controllersadoresUtilizados}

\begin{table*}[ht]
	\scriptsize
	\centering
    \caption{List of controllers addressed in the related literature.}
	\label{tab:controllers}
    \rowcolors{1}{lightgray}{lightgray}
    \begin{tabular}{c|p{0.3\textwidth}p{0.3\textwidth}}
    \hline
    \textbf{Paper} & \textbf{Controller (original)}                                 &   \textbf{Contribution} \\ \hline    	  
    \cite{I17}     & OpenDayLight                                     &                        \\ 
    \cite{I28}     & Floodlight v0.9                                  &                        \\ 
    \cite{I31}     & Floodlight e Beacon                              &                        \\ 
    \cite{I32}     & Ryu                                              &                        \\ 
    \cite{I43}     & POX                                              &                        \\ 
    \cite{I44}     & Floodlight                                                & OperationCheckpoint              \\ 
    \cite{I47}     & FloodLight, Beacon, OpenDayLight and HP VAN SDN  &  Application containment mechanism with \textit{sandboxes} over OpenDaylight Hydrogen \\ 
    \cite{I55}     & FloodLight, ONOS, OpenDayLight, Ryu and POX      &                         \\ 
    \cite{I66}     & POX                                              &                        \\ 
    \cite{I70}     & Floodlight v0.9                                  &                        \\ 
    \cite{A1}      & OpenDayLight Helium                              &                        \\ 
    
    \cite{A2}      & NOX, POX, Beacon, Floodlight, MuL, Maestro and Ryu &                        \\ 
    \cite{A3}      & NOX, Beacon and Maestro                            &                        \\ 
    \cite{A6}      & Floodlight, OpenDaylight, POX and Beacon           & Rosemary               \\ 
    \cite{A10}     & Floodlight, OpenDaylight and ONOS &                        \\ 
    \cite{A12}     & NOX 0.9.1                                         & FortNOX    \\ 
    \cite{A13}     & FloodLight                                       & LegoSDN (FloodLight)   \\ 
    \cite{E4}      & POX                                              &                        \\ 
	\cite{S1}      & Ryu 3.24                                         &  security checksheet                      \\ 
	\cite{S15}     & POX 0.1.0                                        &                        \\ 
	\cite{IS1}     & FloodLight, Maestro, OpenDaylight and POX          &  SPHINX                      \\ 
    \cite{IS2}     & FloodLight                                                  & Security Enforcement Kernel (SEK)    \\ 
	\cite{IS3}     & FloodLight, OpenIRIS, OpenDayLight, Beacon, Maestro, NOX, POX and Ryu &   TopoGuard (FloodLight)                     \\ 
	\cite{Nguyen2017} & FloodLight 1.2              & \\ 
    \cite{alharbi2017} &  Ryu (version 3.22), ONOS (version 1.10.0) and
Floodlight (version 1.0) & \\ \hline
  \end{tabular}
\end{table*}


Currently, there exist hundreds of solutions for SDN controllers. Although many of the these  correspond to proprietary distributions, in the literature we have sampled, most  of the papers describe and validate SDN solutions for open source controllers. Table~\ref{tab:controllers}  lists the   controllers that appear  in the selected papers of the literature. The  Table also shows  the original version used of the controller or the contribution, if improvements have been made  to  the original controller. Notice that FloodLight is one of the most often used controllers with no modifications. We observed that the  controllers chosen for modification are implemented in Java programming language. In these  cases, the popularity of Java is a parameter that may indicate the trend with regard to preference. {Table~\ref{tab:Cont} shows the frequency in  which the controllers appear in the papers we analyzed. The only controller based on proprietary software found in our review is {HP VAN SDN}, with a single occurrence. There are three controllers in  the top-five of the list, which are    implemented in Java:  FloodLight in 1\textsuperscript{st} position, OpenDayLight in 3\textsuperscript{rd}, and Beacon in 4\textsuperscript{th}. There are two controllers implemented in Python,  POX and RYU that are in 2\textsuperscript{nd} and 5\textsuperscript{th} places, respectively. Clearly, One may observe that the popularity of the programming language used in the design of the controller may be  a strong indication of  controller choice.

\begin{table*}[h]
	\scriptsize
	\centering
    \caption{Frequency of controllers in the analyzed papers.}
	\label{tab:Cont}
    \rowcolors{1}{lightgray}{lightgray}
    \begin{tabular}{l|p{4cm}c}
    \hline
    \textbf{Controller}  & \textbf{Papers} & \textbf{Amount} \\ \hline
  FloodLight   & \cite{I28}, \cite{I31}, \cite{I44}, \cite{I47}, \cite{I55},  \cite{A2},  \cite{A6},  \cite{A10}, \cite{A13}, \cite{IS1}, \cite{IS2}, \cite{IS3}, \cite{I70}, \cite{Nguyen2017}, \cite{alharbi2017} & 15 \\ 
      POX          & \cite{I43}, \cite{I55}, \cite{I66}, \cite{A2},  \cite{A6},  \cite{E4},  \cite{S15}, \cite{IS1}, \cite{IS3}                                                           & 9 \\ 
      OpenDayLight & \cite{I17}, \cite{I47}, \cite{I55}, \cite{A1},  \cite{A6},  \cite{A10}, \cite{IS1}, \cite{IS3}                                                             & 8 \\ 
      Beacon       & \cite{I31}, \cite{I47},  \cite{A2},  \cite{A3},  \cite{A6}, \cite{IS3}                                                                                                & 6  \\ 
      RYU          & \cite{I32}, \cite{I55},  \cite{A2},  \cite{S1}, \cite{IS3}, \cite{alharbi2017}                                                                                                           & 6  \\ 
      NOX          &  \cite{A2},  \cite{A3},  \cite{A12}, \cite{IS3}                                                                                                                        & 4  \\ 
      Maestro      &  \cite{A2},  \cite{A3},  \cite{IS1}, \cite{IS3}                                                                                                                        & 4  \\ 
      ONOS         & \cite{I55}, \cite{A10}, \cite{alharbi2017}                                                                                                                                                           & 3  \\ 
      MuL          & \cite{A2}                                                                                                                                                                         & 1  \\ 
      HP VAN SDN   & \cite{I47}                                                                                                                                                                        & 1  \\ 
      Rosemary   & \cite{A6}                                                                                                                                                                        & 1  \\ 
      SPHINX     &  \cite{IS1}                                                                                                                                                                     & 1  \\ 
      OpenIRIS     & \cite{IS3}                                                                                                                                                                        & 1  \\ \hline
    
    \end{tabular}
    
\end{table*}

\subsection{Experimental Environments}
\label{sec:confTopol}


This section presents a data compilation of hardware and software used in the experiments of the analyzed studies. The  experimental environments can  be grouped into  three categories : simulated environment (SE), physical environment (PE) and emulated environment (EE). Table~\ref{tab:setup1a} shows  the classification of the environments, as described in each paper regarding the three categories . The network emulator Mininet\footnote{\url{http://mininet.org/}} was the predominant system in all the works that use emulated environments. 


\begin{table}
\centering
\caption{Types of experimental environments adopted in the analyzed papers.}
\label{tab:setup1a}
\rowcolors{1}{lightgray}{lightgray}
\begin{tabular}{c|ccc}
\hline
\textbf{Paper}                       & \textbf{SE}                   & \textbf{PE}                   & \textbf{EE}                   \\ \hline
\cite{I17}         & \checkmark     &                         & \checkmark     \\
\cite{I28}         &                         & \checkmark     &                         \\
\cite{I41}         &                         & \checkmark     &                         \\
\cite{I31}         &                         &                         &                         \\
\cite{I32}         &                         & \checkmark     &                         \\
\cite{I43}         & \checkmark     &                         & \checkmark     \\
\cite{I44}         & \checkmark     &                         & \checkmark     \\
\cite{I47}         & \checkmark     &                         & \checkmark     \\
\cite{I66}         & \checkmark     &                         & \checkmark     \\
\cite{I70}         & \checkmark     &                         &                         \\
\cite{A1}          &                         & \checkmark     &                         \\
\cite{A2}          &                         &                         &                         \\
\cite{A3}          &                         & \checkmark     &                         \\
\cite{A6}          & \checkmark     & \checkmark     & \checkmark     \\
\cite{A12}         &                         & \checkmark &                         \\
                               &                         &                         &                         \\
\cite{E4}          & \checkmark &                         & \checkmark \\
                               &                         &                         &                         \\
\cite{S1}          &                         & \checkmark     &                         \\
\cite{S15}         & \checkmark     & \checkmark     &                         \\
\cite{IS1}         & \checkmark     &                         &                         \\
\cite{IS2}         &                         & \checkmark     &                         \\
\cite{IS3}         & \checkmark     & \checkmark     & \checkmark     \\
\cite{Nguyen2017}  & \checkmark     &                         & \checkmark     \\
\cite{alharbi2017} &                         &                         & \checkmark     \\ \hline
\end{tabular}
\end{table}

Table~\ref{tab:setup2} shows details of  the software configuration of the experimental environments discussed in each study. In a few cases, the authors clearly indicated the use of virtual machines (VMs) and  we check marked those cases where the authors did  not state  the name or the version of the VMs. With regard to  the operating systems (OS), we included the name and version described in the experiments. In the switch column, we showed  the number  of switches used in the experiments, followed by the model of the switch (virtual/software or physical). The last column shows the number of hosts used in the experiments. All the blank cells refer to studies in which the related item was not described. It should be noted  that \cite{A10,A13,Nguyen2017} did not specify any  of the software used in their experiments.

\begin{table*}[h]
	\scriptsize
	\centering
    \caption{Configurations of software, nodes and data links of the experiments discussed in the analyzed papers.}
	\label{tab:setup2}
    \rowcolors{1}{lightgray}{lightgray}
    \begin{tabular}{c|cc>{\centering}p{0.2\textwidth}c}
    \hline
    \textbf{Study}           & \textbf{VM}          & \textbf{OS}        & \textbf{Switch}                & \# \textbf{Hosts} \\ \hline 
    \cite{I17}      & \checkmark            & Ubuntu 13.04        & 6               & 20          \\ 
    \cite{I28}      &  -                     & -                   & 3x NEC PF5240            & 20          \\ 
    \cite{I41}      &   -                    & -                   & 3x NEC PF5240            & 20          \\ 
    \cite{I31}      &    -                   & -                   & -                        & -           \\ 
    \cite{I32}      &     -                  & -                   & 3x Centec V330           & -           \\ 
    \cite{I43}      &   -                    & -                   & 2               & 6  \\ 
    \cite{I44}      & VirtualBox            & Ubuntu 12.04        & Open Virtual Switch      & -           \\ 
    \cite{I47}      &  -                     & Ubuntu 64-bit       & -                        & -           \\ 
    \cite{I66}      & \checkmark 			& Ubuntu Linux -      & -                        & -           \\ 
    ~               & ~                     & kernel 3.13.0       & -                        & -           \\ 
    ~               &                       & kernel 3.8.0        & HP 5900 (JG336A)  & 2          \\ 

    \cite{I70}      & VirtualBox            & -                   & -                        & 2          \\ 
    ~               &                       & CentOS 6.4          & -                        & -          \\ 
    \cite{A1}       & -		                & Kali Linux          & -                       & -          \\ 
    ~               &                       & Windows 7           & -                        & -          \\ 
    \cite{A2}       &  -                     & Ubuntu 12.04.1      & -                        & -          \\ 
    \cite{A3}       &   -                    & Linux               & HP 5406zl                & -          \\ 
    \cite{A6}       &   -                    & -                   & -                        & 3 \\ 
    \cite{A12}      &   -                    & Ubuntu Server 10.10 & HP ProCurve E6600        & 2          \\ 
    \cite{E4}       & VMWare ESXi           & -                   & -                        & 20         \\ 
    \cite{S1}       & \checkmark            & Ubuntu 14.04        & Pica 8 P-3297            & 3          \\ 
    \cite{S15}      &  -                     & Debian Wheezy       & Open vSwitch 1.10        & 4          \\ 
    \cite{IS1}      &   -                    & Ubuntu 12.04        & 14x IBM RackSwitch G8264 & 10         \\ 
    \cite{IS2}      &    -                   & Linux               & Open vSwitch             & -          \\ 
    \cite{IS3}      &   -                    & -                   & TP-LINK TL-WR1043ND and LINKSYS WRT54G (OpenWRT)   & -          \\ 
\cite{alharbi2017} & \checkmark & - & OpenvSwitch (OVS) version 2.5.0 & -  \\ \hline 
    \end{tabular}
\end{table*}


The overview, obtained from this data compilation, may support the choice for setups of SDN testbeds in future works, as well as providing the  metrics and parameters required for qualitative evaluations between experimental platforms.

\subsection{Main Supporting Tools for the Security Analysis}
\label{sec:tools}


The use of consolidated supporting tools is essential for the experimental  process. As well as  making the process easier and providing greater reliability, consolidated tools can be  cost-effective  and save time employed in the design, implementation and deployment of a specific platform to conduct the experiments.

We enumerated the supporting tools of the literature and estimated  the number of papers citing them. Figure~\ref{fig:tools2} shows  a histogram of the tools applied in security-related experiments in the analyzed literature. The Mininet emulation environment was the most common supporting tool. In the context of experimental evaluation in SDN, Mininet is a general purpose tool. It is able to emulate computer networks through virtualization, by enabling the use of full-featured virtual machines with lightweight operating systems, as well as switches capable of running all the OpenFlow protocols. When instantiating network elements through VMs, Mininet provides flexibility for fast prototyping of realistic network scenarios, with varied topologies and data transmission technologies. Apart from  Mininet, there are two other important supporting tools for packet capturing and network analysis that often appear in the literature: scapy\footnote{\url{http://www.secdev.org/projects/scapy/}} and cbench\footnote{\url{https://github.com/mininet/oflops/tree/master/cbench}}. 

Scapy is a packet manipulation software for capturing, creating, and modifying network packets. On the other hand, cbench is a benchmark tool that is able to configure  a  number of hosts and switches that generate workloads (packet-in) to a target controller, while allowing  its performance to be assessed  through different QoS metrics such as latency and throughput. In the experiments described by the papers we have examined, scapy was used to generate and send a stream  of specific packets or spurious messages to  the network. Otherwise, cbench was generally used to measure the performance of the controllers. We noted  that cbench was applied to determine  the impact in the performance of controllers when a countermeasure is implemented as a new system functionality.

\begin{figure}[h]
\centering
\includegraphics[width=0.9\textwidth]{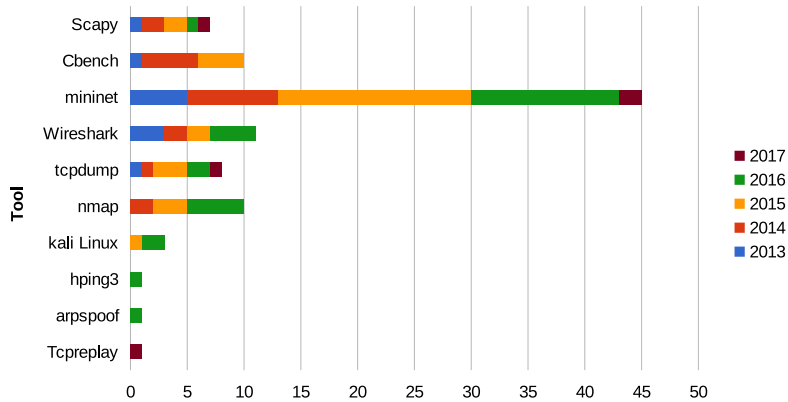}
\caption{List of supporting tools and occurrences by year found in the literature review.}
\label{fig:tools2}
\end{figure}


In addition to scapy and cbench, there are also other common tools used in the literature to perform several types of attacks, as well as for the discovery and exploitation of vulnerabilities in SDNs, such as:

\begin{itemize}
\item \textit{Kali Linux\footnote{\url{http://www.kali.org}}}: an operating system that provides a number of internal tools to find and exploit vulnerabilities in several networked computing systems.
\item \textit{Wireshark\footnote{\url{https://www.wireshark.org}}} and \textit{tcpdump\footnote{\url{https://www.tcpdump.org}}}: sniffer software for capturing and analyzing data packets and network protocols.
\item \textit{Nmap\footnote{\url{https://nmap.org}}}: a software used to perform scanning of open ports on a target remote host.
\item \textit{hping3\footnote{\url{https://tools.kali.org/information-gathering/hping3}}}: a software for launching denial of service (DoS) attacks. 
\item \textit{arpspoof\footnote{\url{http://linux.die.net/man/8/arpspoof}}}: a software for supporting man-in-the-middle (MitM) attacks. 
\item \textit{tcpreplay\footnote{\url{http://tcpreplay.synfin.net/}}}: a software for capturing, modifying and injecting data packets.
\end{itemize}

\section{Security Analysis of Controller Software}
\label{sec:methods}


In this  Section,  we discuss the methods found in the literature dedicated to evaluating the controller software security in SDNs. First, we present a discussion about the methods used for security threat modeling. After this, we conduct a  comparative analysis to show the relationship between the reported studies and security requirements defined by ONF. Finally, we introduce a taxonomy for the existing techniques applied to verify and validate the security requirements of Controller Software in SDNs.

\subsection{Security Threat Modeling and Classification}

In the set of papers, we analyzed the Microsoft  model for security threat classification (STRIDE) ~\cite{P2} and Data Flow Diagrams (DFDs) appear as the most widely used threat modeling techniques.  STRIDE, allows  the security of a system to be classified into six categories: Spoofing, Tampering, Repudiation, Information disclosure, Denial of service, and Elevation of privileges. With the aid of DFDs, the security analyst is able to make   a graphical representation of the software components, the inputs and outputs, as well as the internal  logical  processes \cite{torr2005}.
Besides STRIDE and DFD, there are other alternative methods to assist the modeling of threats and security vulnerabilities. One of them is the Attack Tree, which is able to classify the ways needed to detect  a security breach. The last method, the Petri Net, is a classic mathematical modeling language.




----------------

   


\vrc{ Figure~\ref{fig:polar-metodos} illustrates the trend  in  the use of  threat modeling techniques, as revealed in  the number of published papers, during the years.}

\begin{figure}
\centering
\includegraphics[width=0.6\textwidth]{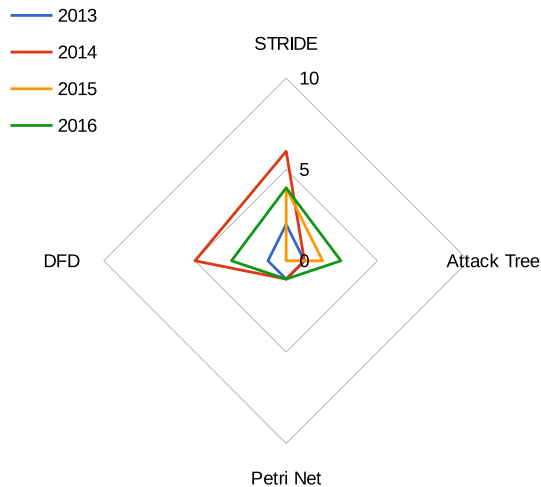}
\caption{Methods for security analysis and the related amount of papers found in our review.}
\label{fig:polar-metodos}
\end{figure}

\vrc{Figure~\ref{fig:polar-stride} shows trends regarding the interest in each STRIDE category, reflected in the number of papers addressing them.} The greatest  interest was displayed in the Denial of Service attack types  since these can  either lead to the network malfunction or collapse. However, there is also an increasing concern about spoofing threats. 

\begin{figure}
\centering
\includegraphics[width=0.8\textwidth]{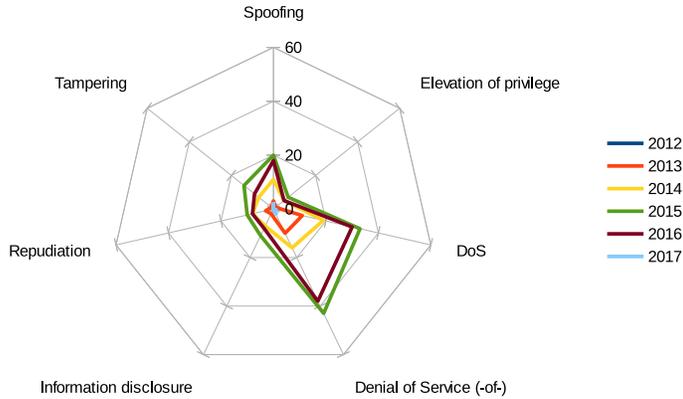}
\caption{The number of papers, found in our literature review, citing each STRIDE category terms over the years.}
\label{fig:polar-stride}
\end{figure}


\vrc{With respect to security properties, Figure~\ref{fig:polar-propriedades} illustrates the trend  in the area of  research, by showing  the number of papers referring each security property}. It should be noted  that the terms ``authentication'' and ``availability'' are mentioned in most  of the papers we analyzed. This is consistent  with the occurrence of the related terms in the STRIDE model. However, it is also clear that authorization  has  also attracted greater attention in the last two years.

\begin{figure}
\centering
\includegraphics[width=0.8\textwidth]{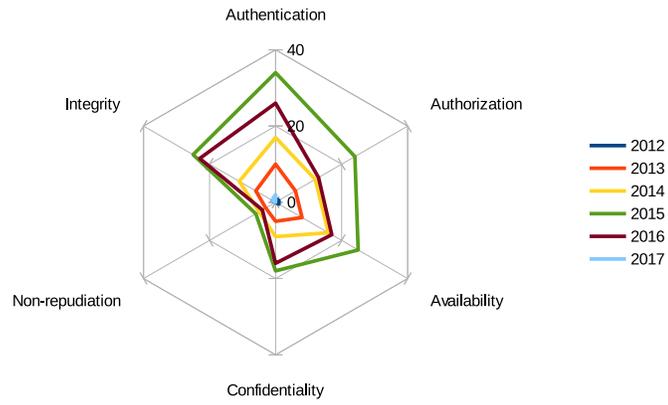}
\caption{The amount of papers, found in our literature review, referring to each security property along the years.}
\label{fig:polar-propriedades}
\end{figure}

According to the review carried out, the STRIDE model is widely adopted and is currently the standard  method for categorizing security threats~\cite{I43}. We  set out guidelines in our literature review with regard to security threats, by  classifying the papers described in Section~\ref{sec:review} based on  the STRIDE model, as shown in Table~\ref{tab:STRIDE}.

\begin{table*}
\scriptsize
\centering
    \caption{Literature studies classified according to the STRIDE model.}
	\label{tab:STRIDE}
    \rowcolors{1}{lightgray}{lightgray}
    \begin{tabular}{c|cccccc}
    \hline
    \textbf{Paper}  & \textbf{S}          & \textbf{T}          & \textbf{R}          & \textbf{I}          & \textbf{D}          & \textbf{E}          \\ \hline
    \cite{I17} &            &            &            &            & \checkmark &            \\ 
    \cite{I32} & \checkmark & \checkmark & ~          & ~          & ~          & ~          \\ 
    \cite{I28} & ~          & ~          & ~          & \checkmark & ~          & ~          \\ 
    \cite{I41} & ~          & ~          & ~          & \checkmark & ~          & ~          \\ 
    \cite{I40} & \checkmark & \checkmark & \checkmark & \checkmark & \checkmark & \checkmark \\ 
    \cite{I43} & \checkmark & \checkmark & \checkmark & \checkmark & \checkmark & \checkmark \\ 
    \cite{I66} & \checkmark & \checkmark & ~          & ~          & ~          & ~          \\ 
    \cite{IS3} & \checkmark & \checkmark & ~          & ~          & ~          & ~          \\ 
    \cite{I55} & \checkmark & \checkmark & ~          & \checkmark & ~          & \checkmark \\ 
    \cite{I31} & ~          & ~          & ~          & ~          & \checkmark & ~          \\ 
    \cite{I44} & ~          & \checkmark & \checkmark & \checkmark & ~          & \checkmark \\ 
    \cite{I47} & ~          & \checkmark & \checkmark & ~          & \checkmark & \checkmark \\ 
    \cite{A1}  & \checkmark & \checkmark & ~          & \checkmark & ~          & \checkmark \\ 
    \cite{A2}  & ~          & ~          & ~          & ~          & ~          & ~         \\ 
\cite{A9}  & \checkmark & \checkmark & \checkmark & \checkmark & \checkmark & \checkmark \\ 
    \cite{A12} & ~          & ~          & ~          & ~          & ~          & \checkmark \\ 
    \cite{A13} & ~          & ~          & ~          & ~          & \checkmark & ~          \\ 
    \cite{E4}  & ~          & ~          & ~          & ~          & \checkmark & ~          \\ 
    \cite{S1}  & \checkmark & \checkmark & \checkmark & \checkmark & \checkmark & \checkmark \\ 
    \cite{IS1} & \checkmark & \checkmark & ~          & ~          & \checkmark & ~          \\ 
    \cite{IS2} & ~          & ~          & ~          & ~          & ~          & \checkmark \\ 
    \cite{IS3} & \checkmark & \checkmark & ~          & \checkmark & \checkmark & ~          \\ 
 	\cite{Nguyen2017}  &  \checkmark &  & ~ &  &  & \\ 
    \cite{alharbi2017} &  &  & ~ &  & \checkmark & \\ \hline   
    \end{tabular}
\end{table*}         

\subsection{ONF's Security Requirements for SDN Controllers}

To the best of our knowledge, the recommendations of ONF ~\cite{TR-529} are the first known security requirements defined for SDN controllers. There are several approaches in the literature that deal with network security in SDN, as  discussed in the previous sections. However, they mostly focus on a single vulnerability or threat. The requirements defined in~\cite{TR-529} are important because they provide the basis of a  security analysis by stressing the need to  evaluate  several points of vulnerability.

\begin{table*}
\scriptsize
\centering
    \caption{Relation between ONF's security requirements and the studies found in the literature review.}
	\label{tab:ONF}
    \rowcolors{1}{lightgray}{lightgray}
    \begin{tabular}{c|p{5cm}lc}
    \hline

 \#    & \textbf{General Sec. Requirements}        & \textbf{Studies}                                                                                & \textbf{Amount}            \\ \hline 
    1    & IP check                          & \cite{I40}, \cite{I43}                                    & 2            \\ 
    2    & User authentication               & \cite{I40}                                                & 1            \\ 
    3    & Account management                & \cite{I40}                                                & 1            \\ 
    4    & Hardware consistency              & -                                                       & 0          \\ 
    5    & Hypervisor security               & -                                                       & 0          \\ 
    6    & Software package integrity        & -                                                  		 & 0            \\ 
    7    & Protecting the integrity of data in transit   & \cite{I32}, \cite{I40}, \cite{I66},\cite{IS3}, \cite{A1}      & 5            \\   
    8.a  & Log function                  & -                                                                             & 0            \\ 
    8.b  & Log files access protection                  & -                                                              & 0            \\        
    8.c  & Log modification protection           & -                                                                     & 0            \\ 
    9    & Protecting the confidentiality          & \cite{I32}, \cite{I40}, \cite{I66},\cite{IS3}, \cite{A1}            & 5            \\ 
    10   & Hiding password and keys display 			 & -                                                             & 0            \\ 
    11   & Traffic separation                  & \cite{I40}, \cite{I43}, \cite{S15}                             & 3            \\ 
    12   & Access control on the GUI     & -                                                                    & 0            \\ 
    13   & VM security          & -                                                                           & 0          \\ 
    14   & Physical host security              & -                                                            & 0          \\ 
    15.a & Anti-DoS from computing Capacity exhaustion   & \cite{I40} e \cite{I31}, \cite{IS1}, \cite{alharbi2017}       & 4            \\ 
    15.b & Closing unnecessary ports/services          & \cite{I40}, \cite{I43}                                          & 2            \\ 
    16   & Authorization for using system functionalities                & \cite{I40}, \cite{I43}, \cite{I47}            & 3            \\ 
    17   & Interface authorization for third parties               & \cite{I40}, \cite{I55}, \cite{I47} , \cite{IS2}     & 4            \\ 
    18   & Security of the hosting OS  				         &  -                                                       & 0         \\ \hhline{====} 

   \#    & \textbf{Specific Sec. Requirements}   & \textbf{Studies}                                                     & \textbf{Amount}            \\ \hline 
    1    & Authentication on interfaces of SDN controllers           & \cite{I40}, \cite{I43}, \cite{E4}, \cite{IS3}             & 4            \\ 
    2    & Protecting reference data from unauthorized modification        & \cite{I40}, \cite{I55}                              & 2            \\ 
    3    & Authorization for access to sensitive data                & \cite{I40}, \cite{I55}, \cite{A1} e \cite{A6}             & 4            \\  
    4    & Hiding password and keys display &  -                                                                                  & 0            \\ 
    5    & Application isolation              & \cite{A12} , \cite{IS2}                                                          & 2            \\ 
    6.a  & Restriction for forwarding packets from switches         & \cite{I28}, \cite{I41}, \cite{I43}, \cite{I40}, \cite{A3} & 5            \\ 
    6.b  & Authorization for flow table creation    & \cite{I40},  \cite{I55}, \cite{I47}, \cite{S15}, \cite{IS2}                & 4            \\  
    6.c  & Anti-DoS from north-/south-bound interfaces (A-CPI, D-CPI)   & \cite{I40}, \cite{I43}, \cite{E4}, \cite{IS1}   	     & 4           \\ 
    6.d  & Anti-DoS from excessive resource consumption         & \cite{I40}, \cite{I31}, \cite{alharbi2017}                    & 3            \\ 
    7    & Privileged control of applications   & \cite{I40}, \cite{I47} , \cite{IS2}                                           & 3            \\ 
    8    & Policy conflict resolution    & \cite{I47}, \cite{A12} , \cite{IS2}                                              & 3            \\ \hline 
    \end{tabular}
\end{table*}

Several surveys have been carried out with focus on  the analysis of security in SDNs \cite{shu2016}, \cite{I60}, \cite{E5}, \cite{A2}, \cite{I63}. However, the  ONF~\cite{TR-529} report  is a technical document containing a comprehensive practical coverage of  security requirements for SDN controllers. 

Although a security evaluation of SND controllers is presented in~\cite{TR-529}, it is restricted to the analysis of five controllers for  a subset of the ONF requirements. In light of this, there are open research issues for further investigation on the security of SDN controllers considering two scopes: in breadth, given that there is a  wide diversity of controllers; and in depth, by taking into account all the ONF requirements.

Table~\ref{tab:ONF} lists the analyzed studies which address both the  general and specific ONF  security requirements~\cite{TR-529}. On  the one hand, it was noted  that some requirements have attracted more interest among researchers, (in Table~\ref{tab:ONF} items $7$, $9$ of General Sec. Req. and items $6.a$ and $6.b$ of Specific Sec. Req.). On the other hand, several requirements were not addressed by any study (in Table~\ref{tab:ONF} items $4$, $5$, $6$, $8$, $10$, $12$, $13$, $14$ and $18$ of General Sec. Req. and item $4$ of Specific Sec. Req.). Among the papers we examined, only~\cite{I17} does not fit into any ONF requirement. Since none of the requirements directly addresses the security in SDN switches, the study \cite{I17} was not included in Table~\ref{tab:ONF}.


\section{Dicussion}\label{sec:discussion}


In most of the papers we analyzed, there is no concern about the details of the  security evaluation methodology that was employed (e.g. description of techniques, experimental setup, assumptions, validation metrics and description of expected results). This lack of detail often hampers  the reproducibility of the experiments, their  continuation or the ability to compare the  experimental results with those of  other studies. We observed that the  methods employed  are  restricted  to the implementation  of different attacks, which were designed to detect the  vulnerabilities of the SDN controllers. 


In addition, in the literature, there is a lack on the definition of groups and categories for the types of techniques used in the security analysis of SDN controllers. We propose a taxonomy as a contribution to fill this gap, which is  illustrated in Figure~\ref{fig:taxonomy}. This taxonomy classifies the security analysis process into two broad categories: \textbf{active}, when the analysis requires introducing of some action in the observed environment; and \textbf{passive},  when only desired events are passively observed in the environment. In the passive analysis, the capture and analysis of network packets were the most common techniques found in the literature employed to discover the network topology of a target SDN network.

\begin{figure*}
\centering
\includegraphics[width=0.9\textwidth]{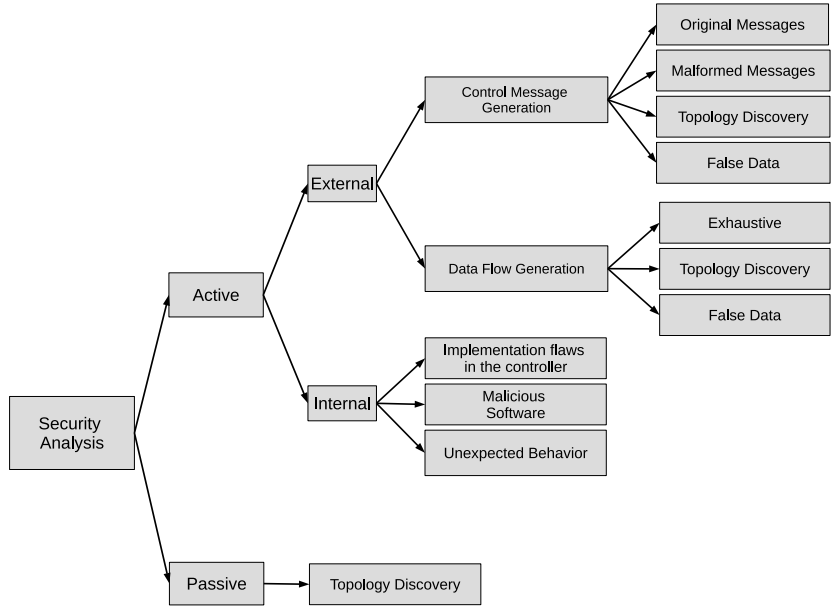}
\caption{Taxonomy considering the techniques used for security analysis of SDN controllers.}
\label{fig:taxonomy}
\end{figure*}


On the basis of  the review we carried out, the active security analysis can be divided into two branches: \textbf{external}, when the action happens is beyond the  scope of the controller, i.e., when it occurs by means of the mediation of other nodes in the network; and \textbf{internal}, when the action takes place  within the controller. 
	

The active-external analysis comprises two techniques: the \textbf{Control message generation} where the  message flow may be either from or to the controller and takes place by means of the (re)injection of original control messages (e.g., Man-in-the-Middle attack), malformed control messages, network topology control messages (e.g., fingerprint) or control messages carrying non-legitimate data (e.g., fake topology); and the \textbf{Data flow Generation} where the flow may be either from or to the controller, and occurs   by means of the injection of an exhaustive data flow (e.g., DoS), a  data flow to discover the  SDN topology (e.g. fingerprint) or data flows conveying illegitimate data (e.g. arp poisoning).


The active-internal analysis can be divided into   three main categories: \textbf{implementation flaws}, \textbf{malicious applications} and  \textbf{unexpected behavior} of the controller. It should be noted that there is a close   correlation between these three subcategories. At first sight  they may seem  similar, since a malicious application may accomplish an attack or an unexpected event  may occur if there is  a failure of implementation (i.e. vulnerability) in the controller. However, the first subcategory is most often related to  flaws in the core of controllers regardless of  the applications. The second, is related to the exploitation of security breaches by malicious applications disregarding any flaw in  the controller. Otherwise, the last subcategory assesses the tolerance of the controller to unintentional threats in which legitimate applications may perform unauthorized/unexpected actions (e.g., an application may exhaust the primary memory, leading to collapse of the controller).

\begin{table*}
\scriptsize
\centering
    \caption{The selected papers in the proposed taxonomy.}
	\label{tab:taxonomy1}
    \rowcolors{1}{lightgray}{lightgray}
    \begin{tabular}{lc}
    \hline
    \textbf{Category}                  & ~ \textbf{Studies}                                                                                         \\ \hline
    --- 1 \textbf{Active}                                          & ~                                                                                       \\ 
    ------ 1.1 \textbf{External}                                     & ~                                                                                    \\ 
    ------------ 1.1.1 \textbf{Control Message Generation}         & ~                                                                                 \\ 
    ------------------ 1.1.1.1 Original Messages                 & \cite{A1}                                                                        \\ 
    ------------------ 1.1.1.2 Malformed Messages                & \cite{A2}                                                           \\ 
    ------------------ 1.1.1.3 Topology Discovery             & \cite{I28}, \cite{I41}                                                         \\ 
    ------------------ 1.1.1.4 False Data                        & \cite{IS1}, \cite{I66}, \cite{Nguyen2017}, \cite{IS3}                                              \\ 
    ------------ 1.1.2 \textbf{Data flow Generation}              & ~                                                                                  \\ 
    --------------- 1.1.2.1 Exhaustive                           & \cite{I43}, \cite{E4}, \cite{S1}, \cite{IS1}, \cite{I17}, \cite{I31}, \cite{alharbi2017} \\ 
    --------------- 1.1.2.2 Topology Discovery              & \cite{I43}, \cite{A3}                                                \\ 
    --------------- 1.1.2.3 False Data                        & \cite{IS1}                                                                       \\ 
    ------ 1.2 \textbf{Internal}                                    & ~                                                                                    \\ 
    ------------ 1.2.1 Imp. flaws in the controller 	& \cite{A6} , \cite{A13}                                                               \\ 
    ------------ 1.2.2 Malicious Software                  	& \cite{I47}, \cite{IS2}, \cite{I44}, \cite{A12}, \cite{I70}, \cite{S15}, \cite{A10}                                                             \\ 
    ------------ 1.2.3 Unexpected Behavior                	& \cite{IS1}, \cite{I32}                                                            \\ 
    --- 2 \textbf{Passive}                                        	& ~                                                                                       \\ 
    ------ 2.1 Topology Discovery                   & \cite{I28}                                                                            \\ \hline
    \end{tabular}
\end{table*}


Table ~\ref{tab:taxonomy1} shows the classification of the approaches described in this survey according to the proposed taxonomy for  the experimental security analysis of the SDN controllers. It is worth noting  that some studies may appear in more than one category, since they perform different  experiments to detect  vulnerabilities in the controllers. It has been found that a  greater effort has been made by  the academic community to conduct  active-external analysis through the exhaustive generation of data flows and the active-internal analysis of Malicious Software. These attempts  reflect a concern that is consistent with the severity of the   security  breach  in those cases, which may lead  to  the controller and also, at some point, the network becoming unavailable.


\subsection{Future Directions}

The literature provides  a range of heterogeneous  techniques that can be  used  although there is a lack  of a common methodology for experimental security analysis, that is applicable to  the overall SDN, and in particular,  the controller software. Such methodology is important to enable  SDN designers to improve the resilience of the  controller core and the resistance of  applications to attacks. 

As the SDN technologies differ on the programming languages and environments, they require some level of abstraction in the definition of this  methodology. However, a methodology for security analysis should  at least describe the  techniques used, the metrics for validation and a list  of expected results. Moreover, it is also preferable to include automatic tests to quantify and certify the security level of the controller software with regard to  the ONF requirements, regardless of the hardware configuration.

\section{Final Remarks}
\label{sec:conclusion}


In this paper we carried out an extensive review of the literature regarding  the security of  controller software in SDNs. We  compiled the main features of the experimental setups and highlighted the many factors involved in the  experimental security evaluation of SDN controllers.


With regard to  the controller software, the literature review showed a tendency of studies related to security issues involving the OpenDayLight, POX and Floodlight controllers. In addition, we enumerated the most common hardware configuration and tools used to  support  the experimental security analysis. Among these tools, the results pointed out a large preference for  the use of Mininet emulator. With regard to  security threat modeling, the review found that the  STRIDE model was widely adopted. This study makes a  contribution to the field by  classifying  literature considering the STRIDE model. Furthermore, we analyzed the relation of the literature to the ONF security requirements and identified the requirements that were met  by each study. Lastly, we created  a taxonomy of the different  approaches found in the literature review considering the techniques employed  used in their experimental security analysis.


It should be stressed  that the literature regarding the experimental security analysis of the implementation  of the network controllers is still limited. Despite the great academic interest in SDN security, there is still a lack of research  on finding a common methodology for automatically evaluating the security of SDN controllers. Additionally, there is  also an absence of methodologies and techniques for automatic risk analysis to estimate the extent of the  damage  that an insecure controller can  inflict on the network.


\bibliographystyle{elsarticle-num}
\bibliography{main}

\begin{thebibliography}{10}
\expandafter\ifx\csname url\endcsname\relax
  \def\url#1{\texttt{#1}}\fi
\expandafter\ifx\csname urlprefix\endcsname\relax\def\urlprefix{URL }\fi
\expandafter\ifx\csname href\endcsname\relax
  \def\href#1#2{#2} \def\path#1{#1}\fi

\bibitem{I16}
S.~Sezer, S.~Scott-Hayward, P.~K. Chouhan, B.~Fraser, D.~Lake, J.~Finnegan,
  N.~Viljoen, M.~Miller, N.~Rao, Are we ready for sdn? implementation
  challenges for software-defined networks, IEEE Communications Magazine 51~(7)
  (2013) 36--43.

\bibitem{I8}
B.~A.~A. Nunes, M.~Mendonca, X.~N. Nguyen, K.~Obraczka, T.~Turletti, A survey
  of software-defined networking: Past, present, and future of programmable
  networks, IEEE Communications Surveys Tutorials 16~(3) (2014) 1617--1634.

\bibitem{I63}
D.~Kreutz, F.~M.~V. Ramos, P.~E. Veríssimo, C.~E. Rothenberg, S.~Azodolmolky,
  S.~Uhlig, Software-defined networking: A comprehensive survey, Proceedings of
  the IEEE 103~(1) (2015) 14--76.

\bibitem{mcKeown2008}
N.~McKeown, T.~Anderson, H.~Balakrishnan, G.~Parulkar, L.~Peterson, J.~Rexford,
  S.~Shenker, J.~Turner, Openflow: Enabling innovation in campus networks,
  SIGCOMM Comput. Commun. Rev. 38~(2) (2008) 69--74.

\bibitem{shu2016}
Z.~Shu, J.~Wan, D.~Li, J.~Lin, A.~V. Vasilakos, M.~Imran, Security in
  software-defined networking: Threats and countermeasures, Mobile Networks and
  Applications 21~(5) (2016) 764--776.

\bibitem{Abdou2018}
A.~{Abdou}, P.~C. {van Oorschot}, T.~{Wan}, Comparative analysis of control
  plane security of sdn and conventional networks, IEEE Communications Surveys
  Tutorials 20~(4) (2018) 3542--3559.

\bibitem{I60}
I.~Ahmad, S.~Namal, M.~Ylianttila, A.~Gurtov, Security in software defined
  networks: A survey, IEEE Communications Surveys Tutorials 17~(4) (2015)
  2317--2346.

\bibitem{dacier2017}
M.~C. Dacier, H.~König, R.~Cwalinski, F.~Kargl, S.~Dietrich, Security
  challenges and opportunities of software-defined networking, IEEE Security
  Privacy 15~(2) (2017) 96--100.

\bibitem{Scott-Hayward2013}
S.~Scott-Hayward, G.~O'Callaghan, S.~Sezer, Sdn security: A survey, in: 2013
  IEEE SDN for Future Networks and Services (SDN4FNS), 2013, pp. 1--7.

\bibitem{I7}
S.~Scott-Hayward, S.~Natarajan, S.~Sezer, A survey of security in software
  defined networks, IEEE Communications Surveys Tutorials 18~(1) (2016)
  623--654.

\bibitem{Li2016}
W.~Li, W.~Meng, L.~F. Kwok, A survey on openflow-based software defined
  networks: Security challenges and countermeasures, Journal of Network and
  Computer Applications 68~(Supplement C) (2016) 126 -- 139.

\bibitem{Dargahi2017}
T.~Dargahi, A.~Caponi, M.~Ambrosin, G.~Bianchi, M.~Conti, A survey on the
  security of stateful sdn data planes, IEEE Communications Surveys Tutorials
  19~(3) (2017) 1701--1725.

\bibitem{TR-529}
I.~Guo, M.~Pourzandi, S.~Scott-Haywar, H.~Song, C.~Wangke, F.~Xialiang, X.~Z.
  Dacheng~Zhang, Security foundation requirements for sdn controllers, Tech.
  Rep. TR-529, ONF (July 2016).

\bibitem{TR-530}
A.~Danping, M.~Pourzandi, S.~Scott-Hayward, H.~Song, M.~Winandy, D.~Zhang,
  Threat analysis for the sdn architecture, Tech. Rep. TR-511, ONF (July 2016).

\bibitem{E5}
A.~Akhunzada, A.~Gani, N.~B. Anuar, A.~Abdelaziz, M.~K. Khan, A.~Hayat, S.~U.
  Khan, Secure and dependable software defined networks, Journal of Network and
  Computer Applications 61 (2016) 199 -- 221.

\bibitem{A2}
A.~Shalimov, D.~Zuikov, D.~Zimarina, V.~Pashkov, R.~Smeliansky, Advanced study
  of sdn/openflow controllers, in: Proceedings of the 9th Central \& Eastern
  European Software Engineering Conference in Russia, CEE-SECR '13, ACM, New
  York, NY, USA, 2013, pp. 1--6.

\bibitem{P2}
S.~Hernan, S.~Lambert, T.~Ostwald, A.~Shostack, Threat modeling-uncover
  security design flaws using the stride approach, MSDN Magazine-Louisville
  (2006) 68--75.

\bibitem{hakiri2014}
A.~Hakiri, A.~Gokhale, P.~Berthou, D.~C. Schmidt, T.~Gayraud, Software-defined
  networking: Challenges and research opportunities for future internet,
  Computer Networks 75 (2014) 453--471.

\bibitem{doria2010}
A.~Doria, J.~H. Salim, R.~Haas, H.~Khosravi, W.~Wang, L.~Dong, R.~Gopal,
  J.~Halpern, \href{https://tools.ietf.org/html/rfc5810}{Forwarding and control
  element separation (forces) protocol specification}~(RFC 5810).
\newline\urlprefix\url{https://tools.ietf.org/html/rfc5810}

\bibitem{pfaff2013}
B.~Pfaff, B.~Davie, \href{https://tools.ietf.org/html/rfc7047}{The open vswitch
  database management protocol}~(RFC 7047).
\newline\urlprefix\url{https://tools.ietf.org/html/rfc7047}

\bibitem{song2013}
H.~Song, Protocol-oblivious forwarding: Unleash the power of sdn through a
  future-proof forwarding plane, in: Proceedings of the second ACM SIGCOMM
  workshop on Hot topics in software defined networking, ACM, 2013, pp.
  127--132.

\bibitem{vasseur2009path}
J.~P. Vasseur, J.~L. Le~Roux, Path computation element (pce) communication
  protocol (pcep), Tech. Rep. RFC 5440 (2009).

\bibitem{enns2006netconf}
R.~Enns, M.~Bjorklund, J.~Schoenwaelder, A.~Bierman, Network configuration
  protocol (netconf), Tech. Rep. RFC 6241 (2011).

\bibitem{bosshart2014p4}
P.~Bosshart, D.~Daly, G.~Gibb, M.~Izzard, N.~McKeown, J.~Rexford,
  C.~Schlesinger, D.~Talayco, A.~Vahdat, G.~Varghese, et~al., P4: Programming
  protocol-independent packet processors, ACM SIGCOMM Computer Communication
  Review 44~(3) (2014) 87--95.

\bibitem{hares2013software}
S.~Hares, R.~White, Software-defined networks and the interface to the routing
  system (i2rs), IEEE Internet Computing 17~(4) (2013) 84--88.

\bibitem{brandt2014security}
M.~Brandt, R.~Khondoker, R.~Marx, K.~Bayarou, Security analysis of software
  defined networking protocols—openflow, of-config and ovsdb, in: The 2014
  IEEE Fifth International Conference on Communications and Electronics (ICCE
  2014), DA NANG, Vietnam, 2014.

\bibitem{I17}
H.~T.~N. Tri, K.~Kim, Assessing the impact of resource attack in software
  defined network, in: 2015 International Conference on Information Networking
  (ICOIN), 2015, pp. 420--425.

\bibitem{I32}
P.-W. Chi, C.-T. Kuo, J.-W. Guo, C.-L. Lei, How to detect a compromised sdn
  switch, in: Proceedings of the 2015 1st IEEE Conference on Network
  Softwarization (NetSoft), 2015, pp. 1--6.

\bibitem{Nguyen2017}
T.-H. Nguyen, M.~Yoo, Analysis of link discovery service attacks in sdn
  controller, in: 2017 International Conference on Information Networking
  (ICOIN), 2017, pp. 259--261.

\bibitem{A13}
B.~Chandrasekaran, T.~Benson, Tolerating sdn application failures with legosdn,
  in: Proceedings of the 13th ACM Workshop on Hot Topics in Networks,
  HotNets-XIII, ACM, New York, NY, USA, 2014, pp. 1--7.

\bibitem{A6}
S.~Shin, Y.~Song, T.~Lee, S.~Lee, J.~Chung, P.~Porras, V.~Yegneswaran, J.~Noh,
  B.~B. Kang, Rosemary: A robust, secure, and high-performance network
  operating system, in: Proceedings of the 2014 ACM SIGSAC Conference on
  Computer and Communications Security, CCS '14, ACM, New York, NY, USA, 2014,
  pp. 78--89.

\bibitem{A10}
S.~Lee, C.~Yoon, S.~Shin, \href{http://doi.acm.org/10.1145/2876019.2876024}{The
  smaller, the shrewder: A simple malicious application can kill an entire sdn
  environment}, in: Proceedings of the 2016 ACM International Workshop on
  Security in Software Defined Networks \&\#38; Network Function
  Virtualization, SDN-NFV Security '16, ACM, New York, NY, USA, 2016, pp.
  23--28.
\newline\urlprefix\url{http://doi.acm.org/10.1145/2876019.2876024}

\bibitem{I31}
F.~Alencar, M.~Santos, M.~Santana, S.~Fernandes, How software aging affects
  sdn: A view on the controllers, in: 2014 Global Information Infrastructure
  and Networking Symposium (GIIS), 2014, pp. 1--6.

\bibitem{I44}
S.~Scott-Hayward, C.~Kane, S.~Sezer, Operationcheckpoint: Sdn application
  control, in: 2014 IEEE 22nd International Conference on Network Protocols,
  2014, pp. 618--623.

\bibitem{I47}
C.~Röpke, T.~Holz, Retaining control over sdn network services, in: 2015
  International Conference and Workshops on Networked Systems (NetSys), 2015,
  pp. 1--5.

\bibitem{A1}
M.~Brooks, B.~Yang, A man-in-the-middle attack against opendaylight sdn
  controller, in: Proceedings of the 4th Annual ACM Conference on Research in
  Information Technology, RIIT '15, ACM, New York, NY, USA, 2015, pp. 45--49.

\bibitem{A12}
P.~Porras, S.~Shin, V.~Yegneswaran, M.~Fong, M.~Tyson, G.~Gu, A security
  enforcement kernel for openflow networks, in: Proceedings of the First
  Workshop on Hot Topics in Software Defined Networks, HotSDN '12, ACM, New
  York, NY, USA, 2012, pp. 121--126.

\bibitem{IS2}
P.~Porras, S.~Cheung, M.~Fong, K.~Skinner, V.~Yegneswaran, {Securing the
  Software-Defined Network Control Layer}, in: Proceedings of the 2015 Network
  and Distributed System Security Symposium (NDSS), 2015.

\bibitem{IS1}
M.~Dhawan, R.~Poddar, K.~Mahajan, V.~Mann, Sphinx: Detecting security attacks
  in software-defined networks., in: Proceedings of the 2015 Network and
  Distributed System Security Symposium (NDSS), 2015.

\bibitem{I28}
R.~Bifulco, H.~Cui, G.~O. Karame, F.~Klaedtke, Fingerprinting software-defined
  networks, in: 2015 IEEE 23rd International Conference on Network Protocols
  (ICNP), 2015, pp. 453--459.

\bibitem{I41}
H.~Cui, G.~O. Karame, F.~Klaedtke, R.~Bifulco, On the fingerprinting of
  software-defined networks, IEEE Transactions on Information Forensics and
  Security 11~(10) (2016) 2160--2173.

\bibitem{A3}
S.~Shin, G.~Gu, Attacking software-defined networks: A first feasibility study,
  in: Proceedings of the Second ACM SIGCOMM Workshop on Hot Topics in Software
  Defined Networking, HotSDN '13, ACM, New York, NY, USA, 2013, pp. 165--166.

\bibitem{I43}
R.~Klöti, V.~Kotronis, P.~Smith, Openflow: A security analysis, in: 2013 21st
  IEEE International Conference on Network Protocols (ICNP), 2013, pp. 1--6.

\bibitem{I66}
T.~Alharbi, M.~Portmann, F.~Pakzad, The (in)security of topology discovery in
  software defined networks, in: 2015 IEEE 40th Conference on Local Computer
  Networks (LCN), 2015, pp. 502--505.

\bibitem{IS3}
S.~Hong, L.~Xu, H.~Wang, G.~Gu, Poisoning network visibility in
  software-defined networks: New attacks and countermeasures., in: Network and
  Distributed System Security (NDSS) Symposium, 2015.

\bibitem{E4}
Q.~Niyaz, W.~Sun, M.~Alam, Impact on sdn powered network services under
  adversarial attacks, Procedia Computer Science 62 (2015) 228 -- 235.

\bibitem{I70}
S.~Betgé-Brezetz, G.~B. Kamga, M.~Tazi, Trust support for sdn controllers and
  virtualized network applications, in: Proceedings of the 2015 1st IEEE
  Conference on Network Softwarization (NetSoft), 2015, pp. 1--5.

\bibitem{S1}
Y.~Hori, S.~Mizoguchi, R.~Miyazaki, A.~Yamada, Y.~Feng, A.~Kubota, K.~Sakurai,
  A Comprehensive Security Analysis Checksheet for OpenFlow Networks, Springer
  International Publishing, Cham, 2017, pp. 231--242.

\bibitem{A11}
D.~Kreutz, F.~M. Ramos, P.~Verissimo, Towards secure and dependable
  software-defined networks, in: Proceedings of the Second ACM SIGCOMM Workshop
  on Hot Topics in Software Defined Networking, HotSDN '13, ACM, New York, NY,
  USA, 2013, pp. 55--60.

\bibitem{S15}
V.~T. Costa, L.~H. M.~K.~Costa, Vulnerabilities and solutions for isolation in
  flowvisor-based virtual network environments, Journal of Internet Services
  and Applications 6~(1) (2015) 18.

\bibitem{alharbi2017}
T.~Alharbi, S.~Layeghy, M.~Portmann, Experimental evaluation of the impact of
  dos attacks in sdn, in: 2017 27th International Telecommunication Networks
  and Applications Conference (ITNAC), IEEE, 2017, pp. 1--6.

\bibitem{Yan2016}
Q.~Yan, F.~R. Yu, Q.~Gong, J.~Li, Software-defined networking (sdn) and
  distributed denial of service (ddos) attacks in cloud computing environments:
  A survey, some research issues, and challenges, IEEE Communications Surveys
  Tutorials 18~(1) (2016) 602--622.

\bibitem{I55}
Y.~E. Oktian, S.~Lee, H.~Lee, J.~Lam, Secure your northbound sdn api, in: 2015
  Seventh International Conference on Ubiquitous and Future Networks, 2015, pp.
  919--920.

\bibitem{torr2005}
P.~Torr, Demystifying the threat modeling process, IEEE Security \& Privacy
  3~(5) (2005) 66--70.

\bibitem{I40}
L.~Yao, P.~Dong, T.~Zheng, H.~Zhang, X.~Du, M.~Guizani, Network security
  analyzing and modeling based on petri net and attack tree for sdn, in: 2016
  International Conference on Computing, Networking and Communications (ICNC),
  2016, pp. 1--5.

\bibitem{A9}
D.~Klingel, R.~Khondoker, R.~Marx, K.~Bayarou, Security analysis of software
  defined networking architectures: Pce, 4d and sane, in: Proceedings of the
  AINTEC 2014 on Asian Internet Engineering Conference, AINTEC '14, ACM, New
  York, NY, USA, 2014, pp. 15--22.

\end{thebibliography}

\end{document}